\documentclass[aps,prx,reprint]{revtex4-1}
\usepackage{lmodern}
\usepackage{graphicx}
\usepackage{braket}
\usepackage{amsfonts, amsmath, amsthm, amssymb}
\usepackage{bbold}
\usepackage{subfigure}
\usepackage{hyperref}
\usepackage{lineno}
\usepackage[normalem]{ulem}
\hypersetup{colorlinks=true,linkcolor=blue,citecolor=blue,urlcolor=blue}
\def\equationautorefname~#1\null{(#1)\null}
\usepackage{cleveref}
\usepackage{braket}
\usepackage[utf8]{inputenc}
\usepackage[T1]{fontenc}
\Crefname{figure}{Fig.}{Figs.}
\newcommand*{\eqnref}[1]{%
	\begingroup
	Eq. \autoref{#1}%
	\endgroup}
\makeatletter
\newsavebox{\@brx}
\newcommand{\llangle}[1][]{\savebox{\@brx}{\(\m@th{#1\langle}\)}%
	\mathopen{\copy\@brx\kern-0.5\wd\@brx\usebox{\@brx}}}
\newcommand{\rrangle}[1][]{\savebox{\@brx}{\(\m@th{#1\rangle}\)}%
	\mathclose{\copy\@brx\kern-0.5\wd\@brx\usebox{\@brx}}}
\makeatother

\begin{document}

\title{Optical conductivity and orbital magnetization of Floquet vortex states}
	
\author{Iman Ahmadabadi}
\author{Hossein Dehghani}
\author{Mohammad Hafezi}
\affiliation{Joint Quantum Institute, NIST and University of Maryland, College Park, Maryland 20742, USA}

%\date{\today}

\begin{abstract}

Motivated by recent experimental demonstrations of Floquet topological insulators, there have been several theoretical proposals for using structured light, either spatial or spectral, to create other properties such as flat band and vortex states. In particular, the generation of vortex states in a massive Dirac fermion insulator irradiated by light carrying nonzero orbital angular momentum (OAM) has been proposed [Kim et al. Phys. Rev. B 105, L081301(2022)]. Here, we evaluate the orbital magnetization and  optical conductivity as physical observables for such a system. We show that the OAM of light induces nonzero orbital magnetization and current density. The orbital magnetization density increases linearly as a function of OAM degree. In certain regimes, we find that orbital magnetization density is independent of the system size, width, and Rabi frequency of light. It is shown that the orbital magnetization arising from our Floquet theory is large and can be probed by magnetometry measurements. Furthermore, we study the optical conductivity for various types of electron transitions between different states such as vortex, edge, and bulk that are present in the system. Based on conductance frequency peaks, a scheme for the detection of vortex states is proposed.

\end{abstract}

% \altaffiliation{Contributed equally}
\pacs{}
\maketitle

\section{Introduction}

A class of condensed matter systems which have gained much attraction in recent years are periodically driven materials, known as Floquet systems, that have resulted a new paradigm for realizing exotic quantum phases of matter \cite{PhysRevBOkia,PhysRevB2010Kitagawa, lindner2011floquet, PhysRevX2014Goldman,Dehghani2014Dissipative,PhysRevB2019Sato,PhysRevLett2018Dejean,PhysRevX2016Paraj}, and some of them have been experimentally realized via optical tools in the last few years \cite{wang2013observation, mahmood2016selective, mciver2020light}. Furthermore, there have been recent experimental developments in spatial manipulation of optical beams for controlling ultra atomic systems  \cite{Zupancic:16,Barredo1021,bernien2017probing,schine2019electromagnetic,yan2014high,PhysRevLett2014Gariepy}. Potentially, applying similar techniques to electronic systems can yield new possibilities for engineering novel quantum phases of matter. In particular, in a recent work \cite{PhysRevBHwanmun}, it was shown that linearly (LP) or circularly polarized (CP) light with nonzero orbital angular momentum (OAM) \cite{Yao2011}, can create vortex states in a two dimensional semiconductor.

More generally, it is intriguing to investigate whether the application of structured light, spatial \cite{PhysRevLettKatan,kim2020optical,PhysRevBHwanmun} or spectral \cite{castro2022floquet},  can lead to interesting topological features, and which physical observables could reveal the properties of bulk, edge, and vortex states in such driven topological systems. 
For example, the frequency-dependence of the optical conductivity provides valuable information about charge carriers and elementary excitations in the dynamical responses. In particular, the real part of the dynamic Hall conductivity describes the reactive carrier response dynamics, and its imaginary part provides the dissipative response \cite{PhysRevLett2010Maciejko,Wu1124,okada2016terahertz,PhysRevBDehghani2015,PhysRevB2015Dehghani,PhysRevB2016Dehghani,PhysRevBDehghani2016,PhysRevResearch2020Nuske,PhysRevLettMak2008,PhysRevLettMorimoto2009,PhysRevLettIkebe2010,qi2009inducing,rokaj2021polaritonic,PhysRevResearch2019Topp,PhysRevLett2014Torres}. Additionally, orbital magnetization, defined by the magnetization arising from orbital motion of electrons, and its origin can yield insightful picture about the electronic properties of system \cite{pershoguba2021optical,PhysRevLettNathan,PhysRevLett2015Dahlhaus,topp2022orbital,PhysRevB2008Souza,PhysRevB2016Bianco,PhysRevLett2005Thonhauser,PhysRevB2020Hara,PhysRevLett2013Bianco,PhysRevB2021Bostr}.

\begin{figure*}[ht]
	\centering
	\includegraphics[width=.8\linewidth]{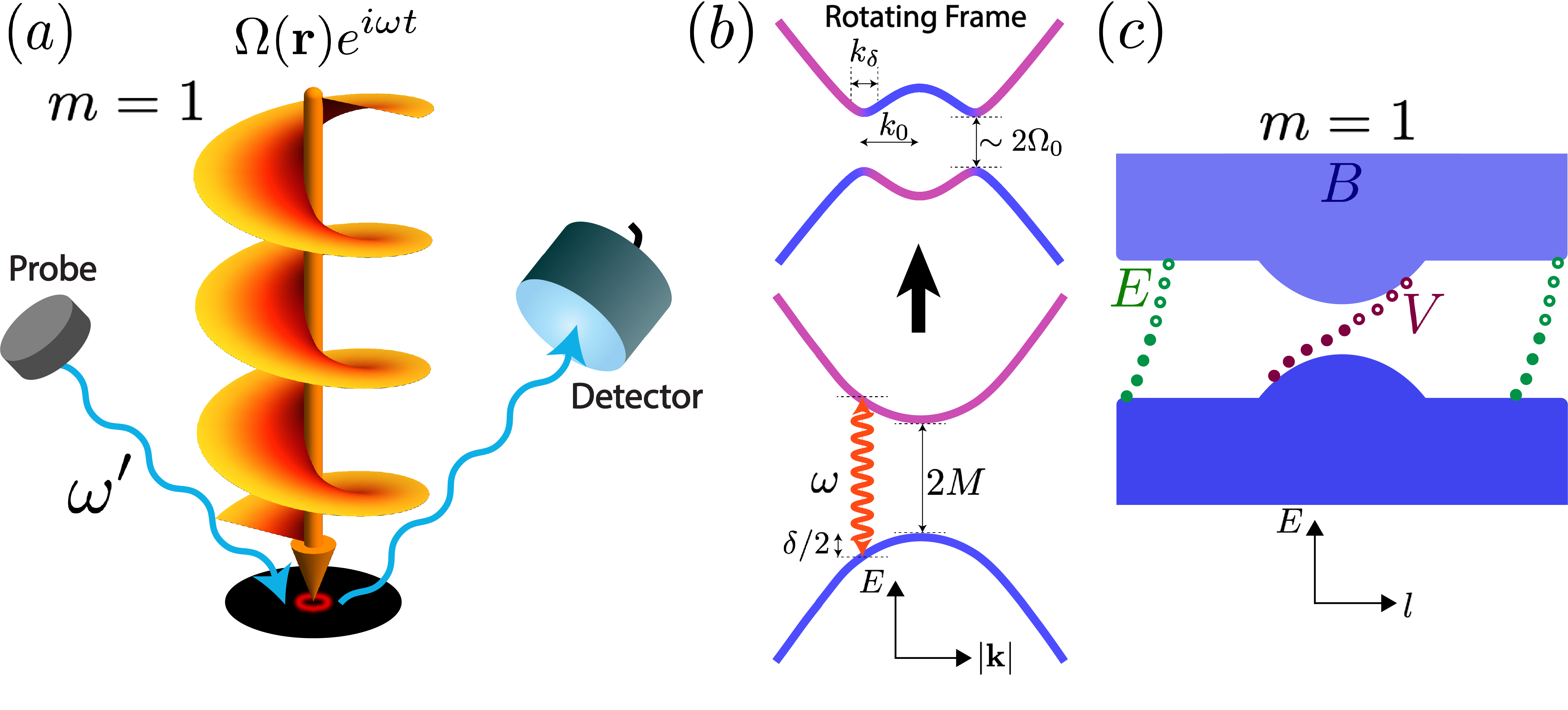}
	\caption{(a) Schematic for the driven 2D semiconductor illuminated by light carrying OAM with vorticity $m=1$. The light beam's frequency is $\omega$. This laser field opens up an energy gap in the rotating frame while creating several midgap (vortex) states. The schematic profile of electrons in a vortex state is shown as the red circle around the light beam's center. Additionally, a setup for measurement of longitudinal and Hall conductivity of the driven system using a prob field with frequency $\omega'$ and a detector to measure the conductivity is shown. (b) In the rotating frame, coupling the laser field with the semiconductor with gap $2M$ and detuning $\delta=\omega-2M$, introduces the hybridization radius $k_{0}$ and the Floquet gap $2\Omega_{0}$ with the thickness $k_{\delta}$. (c) The energy dispersion is plotted as a function of pseudo-OAM $l$ as it becomes a good quantum number regarding the vorticity of light. The bulk, vortex, and edge states are shown in blue area, red, and green dots, respectively.}\label{F1}
\end{figure*}

In this work, we evaluate the optical conductivity and orbital magnetization of a semiconductor driven with structured light, as proposed in \cite{PhysRevBHwanmun}.  To study the Hall conductivity, we follow a finite lattice derivation of dynamical conductivities via Kubo formalism \cite{mahan2013many,PhysRevLettTse,PhysRevBLee,PhysRevLett2010Maciejko}, and we separate the contributions of different types of transitions determined by their initial and final states. This allows us to propose an experimental scheme to measure different contributions to the optical conductivity and to find the experimental signatures of vortex states as shown schematically in \Cref{F1}. Specifically, by tuning the chemical potential and the probe field frequency, we can measure the conductivity contribution arising only from the transition between two vortex states. By further tuning the frequency of the probe field, we can also measure other possible contributions from transitions between bulk, edge, and vortex states \Cref{F1}(c). Since the gap is set by the Rabi frequency, $\Omega_{0}$, the relevant energy scale for bulk-bulk transition is $\Omega_{0}$ while the energy difference between vortex-vortex and edge-edge is given by the light width and system size, respectively. For example, among different types of transitions, vortex-vortex and edge-edge transitions occur at lower probe frequencies compared to those required for detection of the bulk-bulk contributions. The vortex-bulk and edge-bulk transitions need probe frequencies between these two frequency regimes.

Moreover, we study the orbital magnetization and current density of our Floquet system. We show that the orbital magnetization increases linearly with OAM of light, for both LP and CP light. Furthermore, we illustrate that orbital magnetization density is independent of the Rabi frequency and light width. Finally, we demonstrate that in our setting the orbital magnetization density is an intensive quantity because we assume that the light profile covers the entire system.

The orbital magnetization and current density induced by driving the system can be detected based on sensitive magnetometers such as superconducting quantum interference devices (SQUIDs) \cite{AnnalsSQUIDS2022Persky} and nitrogen-vacancy (NV) centers \cite{hong2013nanoscale,glenn2018high,thiel2019probing,sun2021magnetic}. We find that the CP light can create a rotating current density around the center of the light beam in a vortex state, while the density of current is localized along with the polarization of the LP light.

In section \ref{review}, we review the theoretical background of our driven model. Next, in section \ref{dyncond}, we review the Kubo formalism of dynamical conductivities for finite-size systems. We present the results for Hall and longitudinal conductivities as a function of the probe field frequency and different light vorticities and polarizations. In section \ref{OrbMagSec}, the orbital magnetization and current density are introduced for LP and CP light beams. Section \ref{conclusion} presents a discussion of the outlook for future research directions.

\section{Electronic Floquet vortex states of the driven system}\label{review}
In this section, we review our driven system Hamiltonian and the energy spectrum. It is shown that shining light carrying nonzero OAM on a two dimensional semiconductor results in vortex states near resonant and weak field regime \cite{PhysRevBHwanmun}. Specifically, by considering a spinless massive Dirac 2D semiconductor described by the Hamiltonian $H_0 = (vk_x, vk_y, M)\cdot\sigma$, the rotating wave approximation (RWA) for the driven system by light with OAM can be applied. We set $\hbar=1$ and $e=1$ for all calculations, except when these parameters are explicitly determined. Here, $M$ is half of the band gap and $v$ is the Fermi velocity. A light beam with frequency $\omega$ which carries a nonzero OAM is shined on a semiconductor slab, as depicted in \Cref{F1}(a), and its vector potential is denoted by  $\mathbf{\mathcal{A}}(\mathbf{r},t)= A_{0}(r)e^{im\phi}e^{i\omega t}\mathbf{\hat{x}}+\text{c.c.}$. We assume that the laser field satisfies the paraxial approximation, meaning that $A_0(r)=A_\text{max}\left[1-\exp\lbrace-r^2/(2\xi^2)\rbrace\right]$ varies smoothly over the length scale of light width $\xi$. The radial part $A_0(r) e^{im\phi}$, where $r=\sqrt{x^2+y^2}$ and $\phi=\arctan(y/x)$, has integer vorticity $m$, representing the OAM of the laser field. Because of the vortex structure of the field, $A_{0}(r)$ vanishes at $r=0$, for nonzero values of $m$. The applied laser field hybridizes the valence and the conduction bands, opening the energy gap around the resonance ring of momentum, $|\mathbf{k}|=k_0= v^{-1}\sqrt{\omega^2/4-M^2}$ where $\omega>2M$ within the small detuning regime $\delta = \omega - 2M \ll \omega$. Starting with the minimal coupling, one can replace the wave vector $\mathbf{k}$ with $\mathbf{k} + e\mathbf{\mathcal{A}}(\mathbf{r},t)$. To obtain the Floquet Hamiltonian from the time-periodic form $H(t) = H_0 + ev\mathbf{\mathcal{A}}(\mathbf{r},t)\cdot\sigma$, we can use the RWA where we neglect fast oscillating terms in the time dependent Hamiltonian. As it is discussed in the appendix \ref{appA} and shown with more details in reference \cite{PhysRevBHwanmun}, the final form of the Hamiltonian for the LP light reads as follows

\begin{eqnarray}\label{H_rwa1}
H_{\text{RWA}} &=& \frac{\delta}{2}\left(\frac{\mathbf{k}^2}{k_0^2} - 1\right)\sigma_z + \left[ \Omega(r) e^{-im\phi} \sigma_+ + \text{H.c.} \right] \nonumber\\
&& + O\left(\Omega_0\sqrt{\frac{\delta}{M}}\right),
\end{eqnarray}
where $\Omega(\mathbf{r})=evA(\mathbf{r})$ and $\Omega_{0}=\lim_{r\rightarrow\infty}\Omega(\mathbf{r})$. After numerical diagonalization of the Hamiltonian, the dispersion of the Floquet system for the LP light with OAM can be acquired, as depicted in \Cref{F2}(a) for $m=1$. The RWA Hamiltonian for the case of CP is derived in the Appendix \ref{CP_RWA}, and the corresponding energy dispersion is shown in \Cref{F2}(b). As can be observed from \Cref{F2}, there are $|m|$ number of vortex state branches in the energy versus pseudo angular momentum diagram.\\
Based on the formalism followed in previous studies \cite{PhysRevBHwanmun,PhysRevLett2017Abhinav}, here we present an estimation of the energy difference between subsequent vortex states. One can show that the energy separation between vortex states in the low energy regime is given as follows

\begin{eqnarray}
\omega_0 &=& \frac{\int_0^\infty \frac{\Omega(r)}{r} e^{-(2k_0/\delta)\int_0^r \Omega(r') dr'} dr}{k_0\int_0^\infty e^{-(2k_0/\delta)\int_0^r \Omega(r') dr'} dr}.
\end{eqnarray}

We note that this physical quantity is system size-independent and fully determined by the bulk properties of the system and the radial profile of the irradiating beam. Therefore, its value remains the same in the thermodynamic limit. Here, we demonstrate how the energy difference between vortex states can be calculated based on properties of the shining light such as $\omega$, $\delta$, and $\Omega_{0}$. By considering the following form for the radial beam profile

\begin{eqnarray}
\Omega(r) = \left\lbrace
\begin{array}{cc}
\Omega_0 (r/\xi)^q & \text{ for } r\le\xi \\
\Omega_0  & \text{ for } r>\xi
\end{array}
\right.,
\quad q \ge 1.
\end{eqnarray}

Then, it can be shown that $\omega_{0}$ can be approximated by the following expression

\begin{eqnarray}
\omega_0 \simeq & 
\Omega_0 (k_0\xi)^{-1} (k_\delta \xi)^{-(q-1)/(q+1)},
\end{eqnarray}
where, $k_\delta \equiv k_0 \Omega_0 /\delta$. Therefore, the energy separations $\omega_{0}$ depends on the applied light beam properties such as its frequency $\omega$, and the radial profile including parameters $\xi$ for the light width, $\Omega_{0}$ for the intensity, and  $q$ for the light shape. From the approximate energy separation between subsequent vortex states, we can estimate the number of vortex states in one branch as follows

\begin{eqnarray}
2\Omega_0 /\omega_0  \simeq k_0 \xi (k_\delta \xi)^{(q-1)/(q+1)}.
\end{eqnarray}

For $q \ge 1$, one can further simplify this estimation by determining the lower bound of $2\Omega_0 /\omega_0 \simeq   k_0 \xi$. Since we are in the small detuning regime $\xi^{-1} \ll k_{0}\ll M/v$, there are many vortex states in a vortex branch as $2\Omega_0 /\omega_0 \gg 1$.

\begin{figure}[t]
	\centering
	\includegraphics[width=0.9\linewidth]{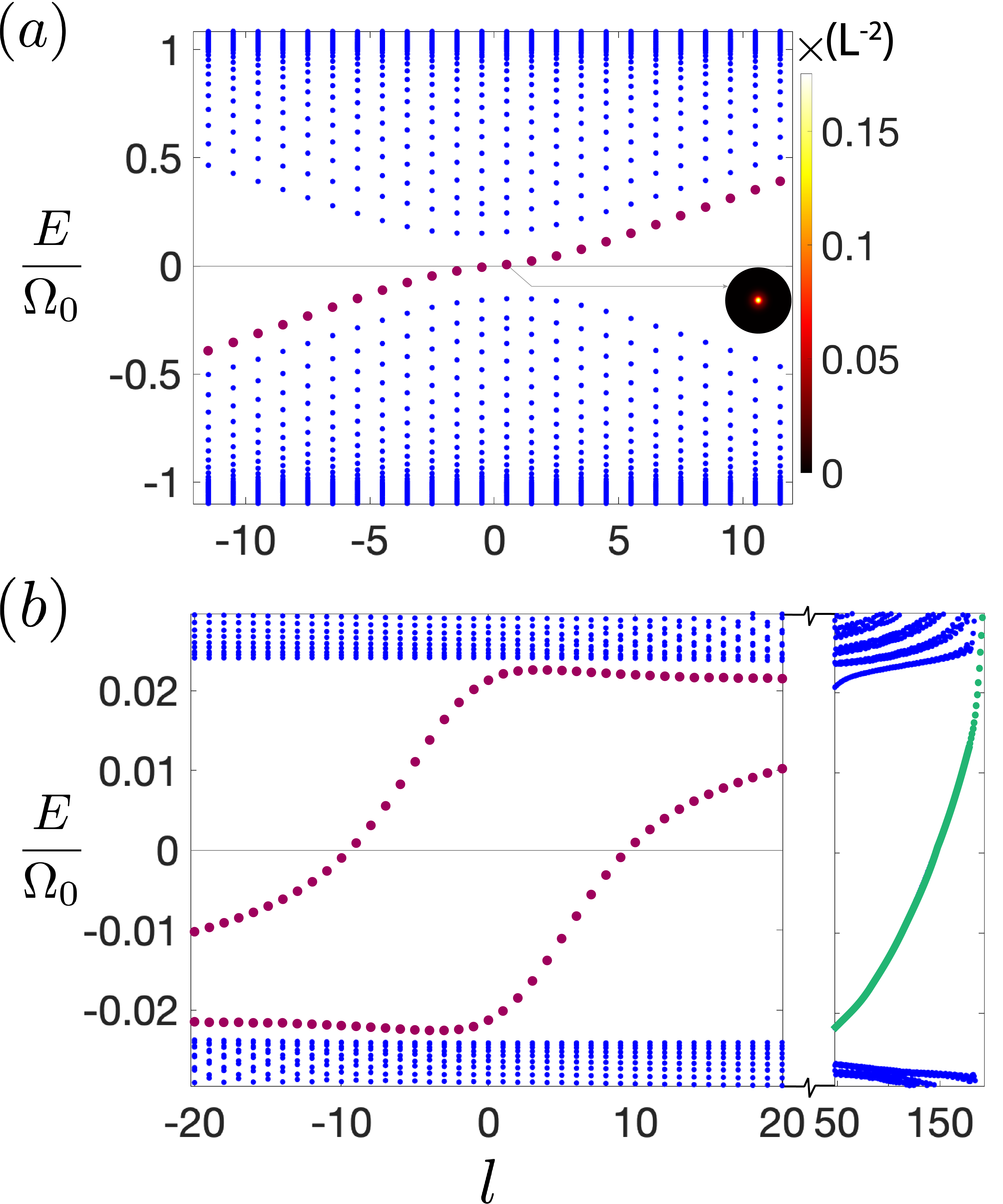}
	\caption{Dispersion of energy versus pseudo-OAM for vorticities (a) $m=1$ of LP light and (b) $m=2$ of CP light. We set $\omega=2.05M$, $A_\text{max}=0.03M (ev)^{-1}$, $A_0(r)=A_\text{max}\left[1-\exp\lbrace-r^2/(2\xi^2)\rbrace\right]$, and $\xi=20k_{\delta}$. We use the disk sample radius $R=10\xi$. As shown in these two examples, the number of vortex state branches can be determined by $|m|$. The bulk and vortex states are bolded in blue and red, respectively. For the CP light in (b), edge states are colored in green. The spatial profile of electronic densities are shown in the inset. Three contributions V-V, V-B, and B-B can be considered for the Hall and longitudinal conductivities of LP light. In addition, two more contributions E-E and E-B are also possible for the CP light.}\label{F2}
\end{figure}

\section{optical conductivity}\label{dyncond}
Here, the optical conductivity of the system described in the previous section is calculated. With the obtained wave functions in the form of Bessel functions and energy spectrum, we can calculate the longitudinal and Hall optical conductivities via Kubo formalism in the real space configuration in polar coordinate. To measure the dynamical conductivity, we apply a weak, linearly polarized AC probe field that is normal to the surface of the semiconductor, as it is shown schematically in Fig.~\ref{F1}(a). Here, we review the real space expression for the dynamical conductivity. We note that, due to the vortex structure of the laser field, the translational symmetry is explicitly broken, and all the calculations of Hall and longitudinal conductivities should be performed on a disc with finite radius $R$. We assume that the non-perturbed Hamiltonian is labeled by $H_{0}$. We consider a time-dependent perturbation $H=H_{0} + H'(t)$ and apply the Liouville-von Neumann equation $i\hbar \partial_{t}\rho = [H,\rho]$ for the density matrix $\rho=\rho_{0} + \delta\rho$ in the linear response regime. Relations $H_{0}\ket{\alpha}=\epsilon_{\alpha}\ket{\alpha}$, $\rho_{0}\ket{\alpha}=n_{\alpha}(\epsilon)\ket{\alpha}$ can be used in Liouville-von Neumann equation, where $n_{\alpha}$ is the Fermi-Dirac distribution. In other words, we assume that the system is thermalized in the rotating frame. With the assumption of sinusoidal time dependence of $H'(t)\varpropto e^{i\omega't}$, we have $\hbar \omega' \delta\rho=[H',\rho_{0}]+[H_{0},\delta\rho]$, where $\omega'$ is the frequency of the prob field. Thus the components of $\delta \rho$ can be obtained as follows
\begin{align}
\braket{\beta|\delta\rho|\alpha} = \dfrac{n_{\beta}-n_{\alpha}}{(\epsilon_{\beta}-\epsilon_{\alpha})-\hbar\omega'-\mathrm{i}\eta}\braket{\beta|H'|\alpha},
\label{eq:drho_el}
\end{align}
where $ \mu,\nu \in \set{x,y} $, $\eta$ is the quasiparticle lifetime broadening, $ \alpha (\beta) $ is a collective label for the relevant quantum state including the band index $n$, pseudo-OAM $l$, and the type of the state from the set $\set{\text{bulk, }\text{vortex, }\text{edge}}$.
\begin{figure*}[ht]
	\centering
	\includegraphics[width=1.00\linewidth]{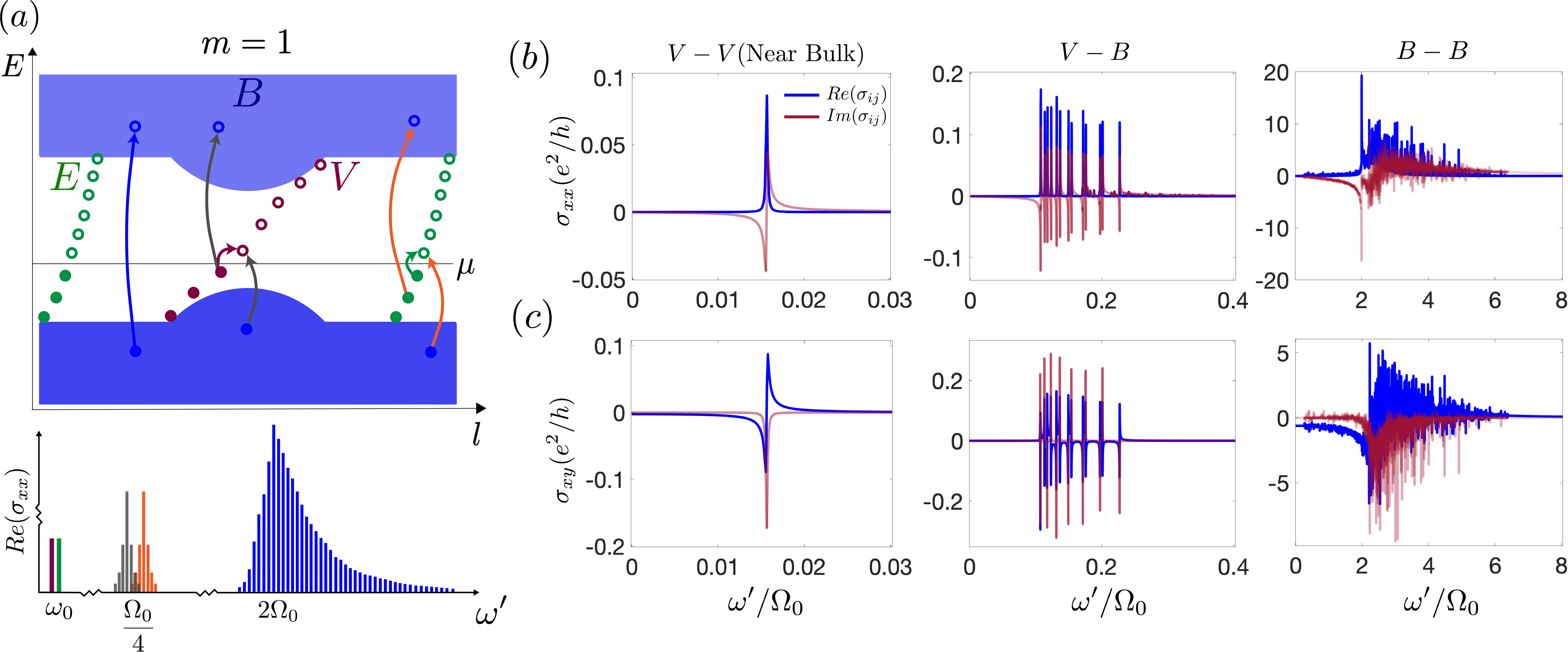}
    	\caption{(a) Schematic of energy dispersion illustrating that tuning the probe field frequency through different regimes can measure different optical conductivity contributions. The bulk, vortex, and edge states are shown in blue areas, red, and green dots, respectively. Transitions between V-V, V-B, E-E, E-B, and B-B are shown in red, gray, green, orange, and blue arrows, respectively. The corresponding location of each type of transition for $Re(\sigma_{xx})$ is shown below schematically, for a system driven by LP light. (b) Numerical results for the longitudinal and (c) the Hall conductivity of LP light $ \sigma_{xy} $ and $ \sigma_{xx} $, versus probe frequency $ \omega'$ for contributions from different types of transitions. Parameters here are the same as shown in \Cref{F2} and the vorticity of light is $m=1$. Blue and red lines correspond to the real and imaginary parts of conductivity, respectively.}\label{F3}
\end{figure*}

\begin{figure*}[ht]
	\centering
	\includegraphics[width=1.00\linewidth]{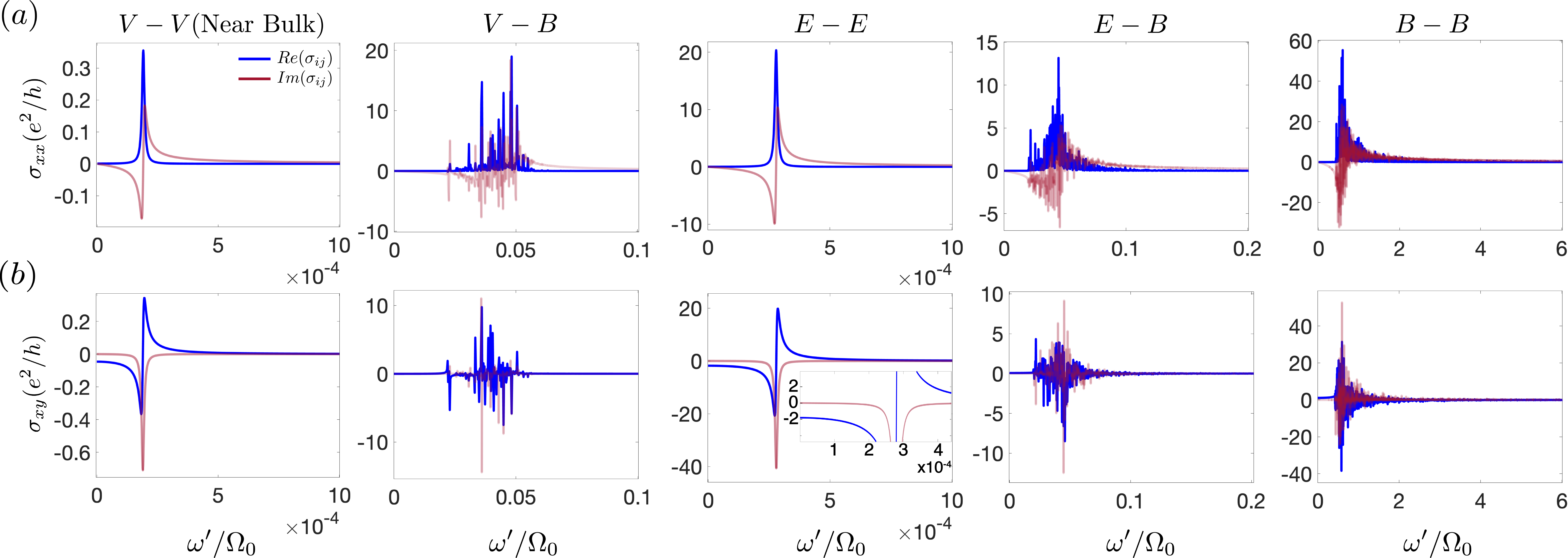}
	\caption{The optical conductivities, Hall ($ \sigma_{xy} $) and longitudinal ($ \sigma_{xx} $), for CP light, versus probe frequency $ \omega'$ and vorticity $m=1$ for different types of transitions. Parameters are the same as in \Cref{F2}. Blue and red lines indicate the real and imaginary parts of optical conductivity, respectively.}\label{F4}
\end{figure*}
We can rewrite the perturbation $H'=eA^{\mu}_{p}p_{\mu}/m=ev_{\mu}E_{\mu}/(i\omega')$, where $e>0$ is the elementary charge and Einstein summation rule is applied. Using Heisenberg equation of motion for time-dependent perturbations, one obtains $ \bra{\beta}v_{\mu}\ket{\alpha} = \bra{\beta}\dot{x}_{\mu}\ket{\alpha} = \bra{\beta}[x_{\mu},H]\ket{\alpha}/(i\hbar)  \approx  \bra{\beta}x_{\mu}\ket{\alpha}(\epsilon_{\alpha} - \epsilon_{\beta})/(i\hbar) $. The single particle current density operator is defined as $j_{\mu}=\frac{1}{\mathcal{A}}\frac{\delta H}{\delta A^{\mu}_{p}(\mathbf{r})}=(-e)v_{\mu}/\mathcal{A}$, where $\mathcal{A}$ is the area of the two dimensional system (here, $\mathcal{A}=\pi R^{2}$), and the average paramagnetic current density can be calculated as $J_{\mu}=Tr[j_{\mu}\delta\rho]$. By substituting the previous relations into the average current density, we get
\begin{align}
J_{\mu} & =\sum_{\alpha\beta}\braket{\beta|\delta\rho|\alpha}\braket{\alpha|j_{\mu}|\beta}    \label{Jpara}  \\ %\nonumber \\ 
&=-\frac{2\pi}{\hbar} \sigma_{0}\sum_{\alpha\beta}\dfrac{(n_{\beta}-n_{\alpha})\epsilon_{\beta\alpha}^2}{\epsilon_{\beta\alpha}-\hbar\omega'-\mathrm{i}\eta}\dfrac{\braket{\alpha|x_a|\beta}\braket{\beta|x_b|\alpha}}{\mathcal{A}}A_{\nu}, \nonumber 
\end{align}
where $ \epsilon_{\beta\alpha}\equiv\epsilon_{\beta}-\epsilon_{\alpha} $ is the energy difference between final ($\beta$) and initial ($\alpha$) transition states and $ \sigma_{0}\equiv e^2/h $ is the quantum of conductance. The paramagnetic current correlation function $\Pi^{\mu\nu} (\omega')$ defined by the equation $J_{\mu} = \Pi^{\mu\nu} (\omega')A_{\nu}$ is as follows
\begin{align}\label{eq:cond_ab}
\Pi^{\mu\nu}(\omega')&=-\frac{2\pi}{\hbar} \sigma_{0}\sum_{\alpha\beta}\dfrac{(n_{\beta}-n_{\alpha})\epsilon_{\beta\alpha}^2}{\epsilon_{\beta\alpha}-\hbar\omega'-\mathrm{i}\eta}\dfrac{\braket{\alpha|x_{\mu}|\beta}\braket{\beta|x_{\nu}|\alpha}}{\mathcal{A}}.
\end{align}
Therefore, the following final equation for the dynamical conductivity can be acquired
\begin{align}\label{opt_cond_final}
\sigma_{\mu\nu}(\omega')= & \frac{\Pi^{\mu\nu}(\omega')-\Pi^{\mu\nu}(0)}{\mathrm{i}\omega'} \nonumber \\
=& \dfrac{2\pi\mathrm{i}}{\mathcal{A}} \sigma_0\sum_{\alpha\beta}\dfrac{(n_{\beta}-n_{\alpha})\tilde{\epsilon}_{\beta\alpha}}{\tilde{\epsilon}_{\beta\alpha}-\omega'-\mathrm{i}\eta}{\braket{\alpha|x_\mu|\beta}\braket{\beta|x_\nu|\alpha}},
\end{align}
where $\tilde{\epsilon}_{\beta\alpha}\equiv\epsilon_{\beta\alpha}/\hbar$. We set $\eta=9.6\times 10^{-6}$ and $\eta=5.6\times 10^{-7}$ for irradiating light carrying OAM with LP and CP, respectively. The reason we need to use different values for $\eta$ is due to the different Hamiltonian energy scales for the LP and CP cases as shown in \eqnref{H_rwa1} and \eqnref{Hrwa_cp}. The matrix elements in \eqnref{opt_cond_final} capture the transition processes among the bulk, edge, and vortex states. There are five types of transitions, \textit{i.e.}, edge-to-edge (E-E), edge-to-bulk (B-E), vortex-to-vortex (V-V), vortex-to-bulk (V-B), and bulk-to-bulk (B-B). In the system, the V-E transition is not possible because vortex and edge states' branches are located in separate ranges of pseudo-OAM $l$ in the dispersion, sufficiently far from each other so that the transition rule, $l'=l \pm 1$,  cannot be obeyed. Using the wave functions' expressions in \eqnref{eigstate}, we obtain the following matrix form with corresponding transition rules for various contributions. We note that the transition rules $l=l'\pm1$ are obtained by integrating the angular part of terms $\braket{\alpha|x_\mu|\beta}$.

\begin{align}\label{matrix_element}
\Braket{\psi_{l'n'}^{S'}|\hat{x}|\psi_{ln}^{S}}
&=\ \mathcal{T}_{l'n',ln}^{S'S}(\delta_{l,l'+1} + \delta_{l,l'-1})\nonumber\\
\Braket{\psi_{l'n'}^{S'}|\hat{y}|\psi_{ln}^{S}}
&=\ \mathcal{T}_{l'n',ln}^{S'S}(i\delta_{l,l'+1} -i\delta_{l,l'-1}),
\end{align}
where  $ \delta_{l,l'} $ is the Kronecker delta symbol, $ S', S\in\set{\text{B}, \text{E}, \text{V}} $ represents the bulk, edge, and vortex states, and $ \mathcal{T}_{l'n',ln}^{S'S} $ is a dimensionless radial integration part derived by

\begin{equation}\label{trans-integral}
    \mathcal{T}_{l'n',ln}^{S'S} = \int_{0}^{R}\mathrm{d}rr^{2} (u_{n,l,+}(r)u_{n',l',+}(r) + u_{n,l,-}(r)u_{n',l',-}(r)).
\end{equation}
In \Cref{F3} and \Cref{F4}, different contributions of the optical conductivity (Hall and longitudinal) for LP and CP light beams are shown, respectively. \Cref{F3}(a) is illustrating the electronic transfer among different types of states that includes V-V, V-B, E-E, E-B, and B-B transitions, which is shown in red, gray, green, orange, and blue, respectively. According to the transition rule obtained in the Kubo formalism in \eqnref{matrix_element}, it can be seen that V-V transition occurs only between a state below and a state above the Fermi level. Thus, there is only one resonance peak in $m=1$ for V-V in \Cref{F3}, corresponding to the energy difference between two subsequent vortex states that can be approximated by $\omega_{0}$ as discussed in section \ref{review}. We also note that since the two vortex states at zero energy are particle-hole symmetric to each other as we showed below \eqnref{heff}, the radial parts of their wave functions cancel each other out. As a result, the intensity of the V-V transition at zero energy vanishes. Therefore, we need to change the chemical potential to select two vortex states far from the zero energy so that the transition between two subsequent vortex states becomes nonzero. Despite the behavior of V-V transitions for $m=1$, there are many peaks in the V-B contributions. In \Cref{F3}, the V-B has several peaks because of several transitions between vortex states in the gap and possible bulk states, corresponding to transition frequencies between these states which for our parameters is around the energy $\omega'\sim \Omega_{0}/4$. The B-B contribution has even more peaks in comparison to the case of V-B since the bulk transitions scale with the system's area. The location of peaks covers energy differences in the range $2\Omega_{0} < \omega'$ and decays for higher probe frequencies.

The V-V transition for the light carrying OAM with CP shown in \Cref{F4} has very similar behavior to the LP light, where the peaks for $\sigma_{xx/xy}$ occur at resonance with the energy difference between two vortex states, $\omega_{0}$. We note that energy scales for CP light are smaller than LP light by a factor of $v^{2}/2M$ as it is demonstrated in the Hamiltonian in \eqnref{Hrwa_cp}.
As can be observed from \Cref{F4}, for the case of CP light, the E-E contribution to the Hall conductivity is dominant. The E-E contribution reaches the value $\sim 1.95$ as $\omega'\rightarrow 0$, corresponding to the quantized Hall conductance, according to the existing two chiral edge modes and Chern number two for topological Floquet insulator. In the E-B contribution of CP light, similar to the case of LP light, there are more possible transitions than the cases of V-V and E-E as shown in \Cref{F4}. The V-B transition for the LP light shown in \Cref{F3} has distinct peaks because of the more separate vortex states (larger $\omega_{0}$) in the gap of the driven system in comparison to the case of CP light. The B-B transition for CP illumination has more resonance peaks than all other transition types. Similarly, the reason is that more possible electron transfers obeying transition rules between bulk states are available in comparison to other contributions. For the CP light, most B-B peaks occur around frequency range $v^{4}\Omega_{0}/2M < \omega' $ and decay exponentially at higher frequencies.
Different types of transitions are also discussed for the OAM of light $m=2$ for the LP light in the Appendix \ref{m2LPHall} and \Cref{F8}. Most of the contributions are very similar to the case of $m=1$, except the V-V transition for vorticity $m=2$ has more peaks for electron transfer between vortex states. This is because, for $m=2$, transitions between vortex branches are also possible and introduce more peaks as it is depicted in \Cref{F8}. 

We note that it is not possible to measure the optical conductivity of vortex states \emph{locally}. This is because the wavelength of the probe field for V-V transition is of the order of $\lambda\sim\frac{1}{\omega_{0}} \gg R$, the system size, and thus larger than the radius of localized electronic density in a vortex state that is located around the center of the light as shown schematically in \Cref{F1}(a).

\begin{figure}[t]
	\centering
	\includegraphics[width=1.00\linewidth]{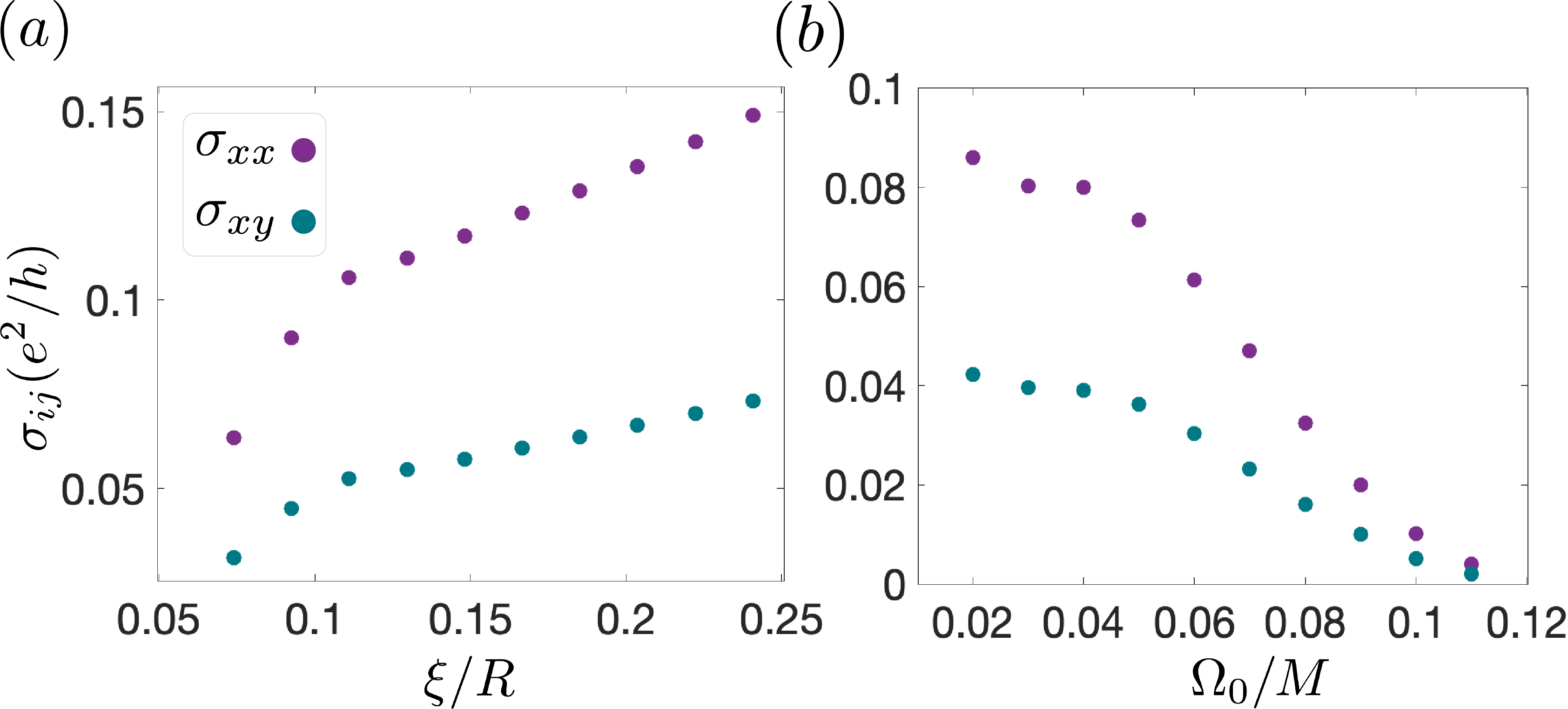}
	\caption{(a) Optical conductivity as a function of light width $\xi$, and (b) as a function of Rabi frequency $\Omega_{0}$ for V-V transition of LP light as it is depicted in \Cref{F3}(c).}
	\label{F10}
\end{figure}

However, we can show that it is possible to distinguish transitions between different types of states spectrally. To separate different contributions of conductivity in experiments, one can use the optical conductivity measurements by tuning the probe field frequency properly. To detect vortex states and measure their contributions to the dynamical conductivity, the chemical potential should be tuned to be in the bulk gap of the driven system and not exactly at energy zero. The reason for the latter condition is because of the vanishing amplitude of the transitions between vortex states above and below the energy zero as their radial integration in \eqnref{trans-integral} vanishes. By tuning the probe field frequency $\omega'$ to be less than the bulk gap, one can remove any bulk contributions as shown by the red transition in \Cref{F3}(a). Then to measure the B-V contribution, the probe frequency can be tuned to include B-V contributions as illustrated in \Cref{F3}(a) in gray     transition. After measuring the contribution of V-V and V-B, by choosing the probe frequency to be equal or higher than the bulk gap, B-B contribution can be possible and measured as shown with blue transition in \Cref{F3}(a). We note that the amplitudes of V-V transitions are system size-dependent and they decrease as the radius of the system, $R$, increases. However, here we use this finite-size effect to acquire the signature of the vortex states in the optical conductivity. We can also tune Rabi frequency $\Omega_{0}$ and light width $\xi$ to change the intensity of optical conductivity. As it is shown in \Cref{F10}, the optical conductivity can increase as a function of light width and decrease when the Rabi frequency increases.

To verify the experimental feasibility of the optical conductivity measurements in our system,  we  note that realizing our Floquet system, similar to other recent studies \cite{mahmood2016selective,mciver2020light}, requires strong laser fields. While so far the experiments have been performed on gapped states, we use their numbers as a guide for our proposal in semiconductors. The intense fields pump a considerable amount of energy into the system, and therefore can quickly heat the system. Therefore, in such settings where Floquet states have been shown to survive for around $1 \mathrm{ps}$, our proposed vortex states can be created transiently. Correspondingly, to measure the physical signatures of these states, one needs to consider an ultrafast measurement protocol. The typical vector potential and detuning that we have assumed in our proposal are $A_{0}=0.015M(ev)^{-1}$ and $\delta = 0.1M$. This value corresponds to the Rabi frequency $\Omega_{0}=evA_{0}=3.6 \mathrm{ THz}$ for a semiconductor band gap $M \sim 1 \mathrm{ eV}$ and Fermi velocity $v\sim 10^{5}\mathrm{ m/s}$. The corresponding intensity for a such a Rabi frequency is $I=\frac{c\epsilon_{0}}{2}\omega^{2}A^{2}=2.1\times 10^{12} \mathrm{ W/m^{2}}$ that is close to the intensity used in Ref. \cite{mciver2020light}, where $c$ is the speed of light and $\epsilon_{0}$ is the dielectric permittivity of vacuum. Based on \Cref{F3} and \Cref{F4}, optical conductivity peaks in our system occur in the range of probe frequencies $\omega'\sim \frac{1}{100}\Omega_{0}$ to $\omega'\sim \Omega_{0}$. Here, the inverse of the probe frequency can be compared with the duration of the recent ultrafast DC measurement of anomalous Hall conductivity in the driven graphene \cite{mciver2020light}. The inverse of probe frequencies in optical conductivity can be within the range $30-1000$ fs and are less than the duration of such experiments. Therefore, we conclude that our measurement scheme for optical conductivity of different types of transitions is experimentally achievable.

\section{orbital magnetization and current density}\label{OrbMagSec}
\begin{figure}[t]
	\centering
	\includegraphics[width=1\linewidth]{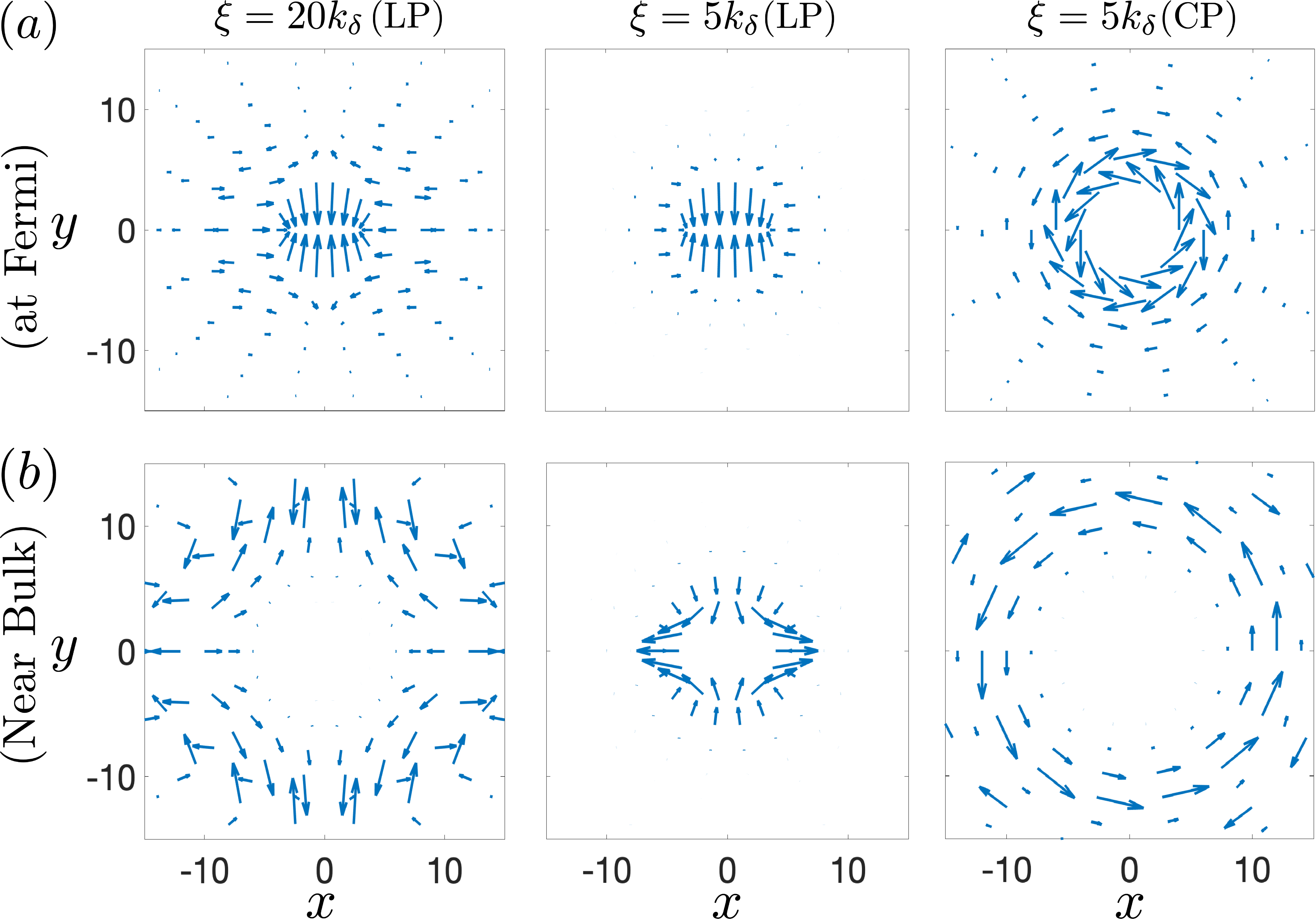}
	\caption{(a) Current density of vortex states, demonstrating highly localized current density around the center of light for the vortex states closest to the zero energy. (b) Current densities for the vortex states far from the zero energy and near the bulk states. Parameters for both (a) and (b) are the same as \Cref{F2}, except $\xi$ that is determined here separately.}\label{F5}
\end{figure}

To further understand the effect of the vorticity of light on the electronic system, we calculate the electronic current density and orbital magnetization. Here, for the wave functions of the quantum states, $\psi(\bf{r}) = \langle\bf{r}|\psi_{m,\bf{k}}\rangle$, the current density is given by 

\begin{equation}
    \bf{j}(\bf{r}) = -e\psi^{\dagger}(\bf{r})\frac{\partial H_{RWA}}{\partial \bf{k}} \psi(\bf{r}).
\end{equation}

\begin{figure}[t]
	\centering
	\includegraphics[width=1.00\linewidth]{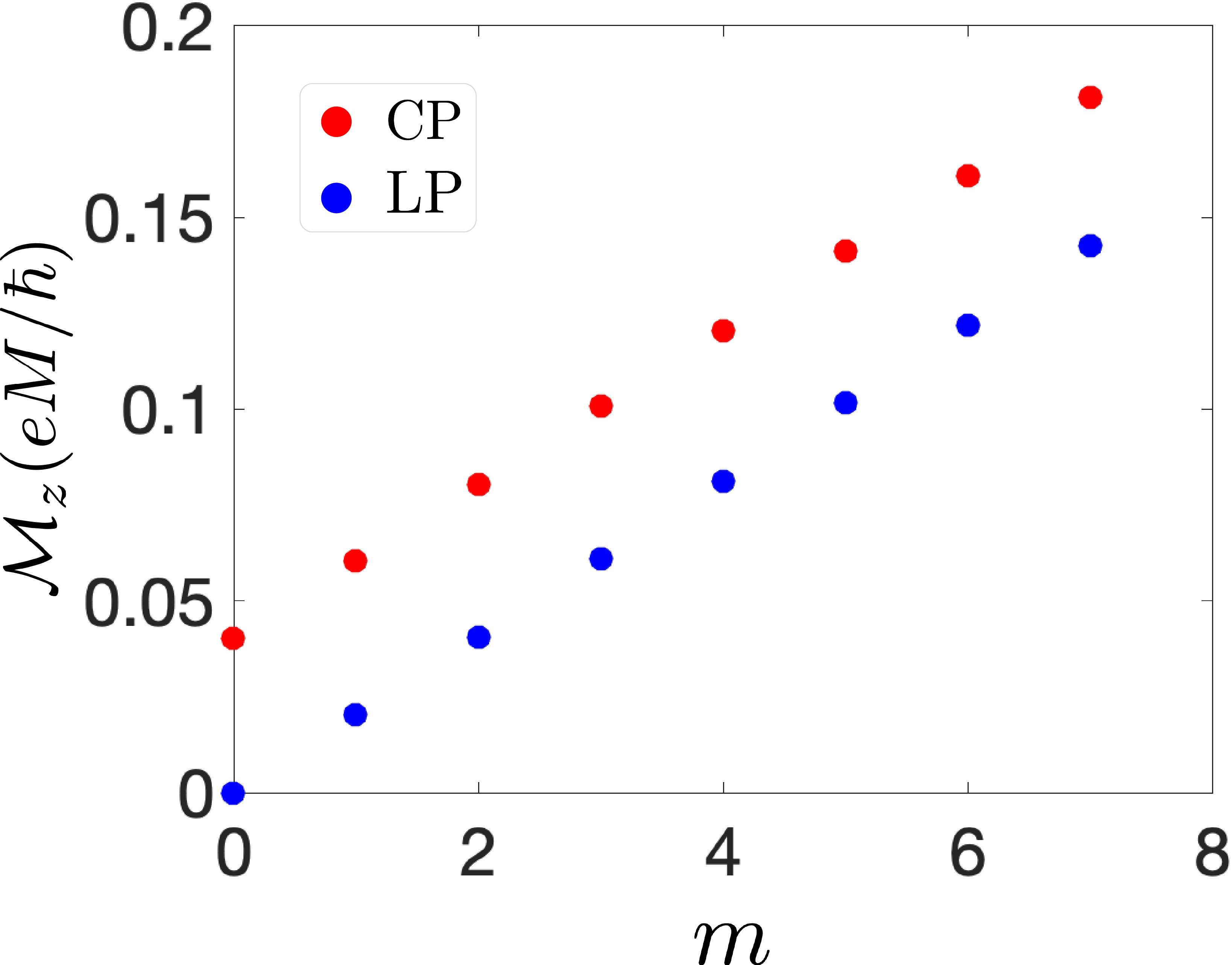}
	\caption{Orbital magnetization of the Floquet system, indicating linear increase in the diagram of orbital magnetization as a function of $m$ for both CP and LP laser fields. Same set of parameters as in \Cref{F2} are used here.}\label{F6}
\end{figure}

As it is shown in Fig.~\ref{F5}(a), the current density of a vortex state is highly localized around the center of the light carrying nonzero OAM for both LP and CP laser fields. In the case of linear polarization, the current density is aligned linearly along the polarization of the light beam. The width of this localization increases as $\xi$ increases and as we select the vortex states far from the zero energy. The rotation of the current density for the CP case is detected by the handedness of the beam. Then, we calculate the orbital magnetization of occupied states in the presence of vortex states for different vorticities $m$. The orbital magnetization is defined as follows 
\begin{figure}[t]
	\centering
	\includegraphics[width=1.00\linewidth]{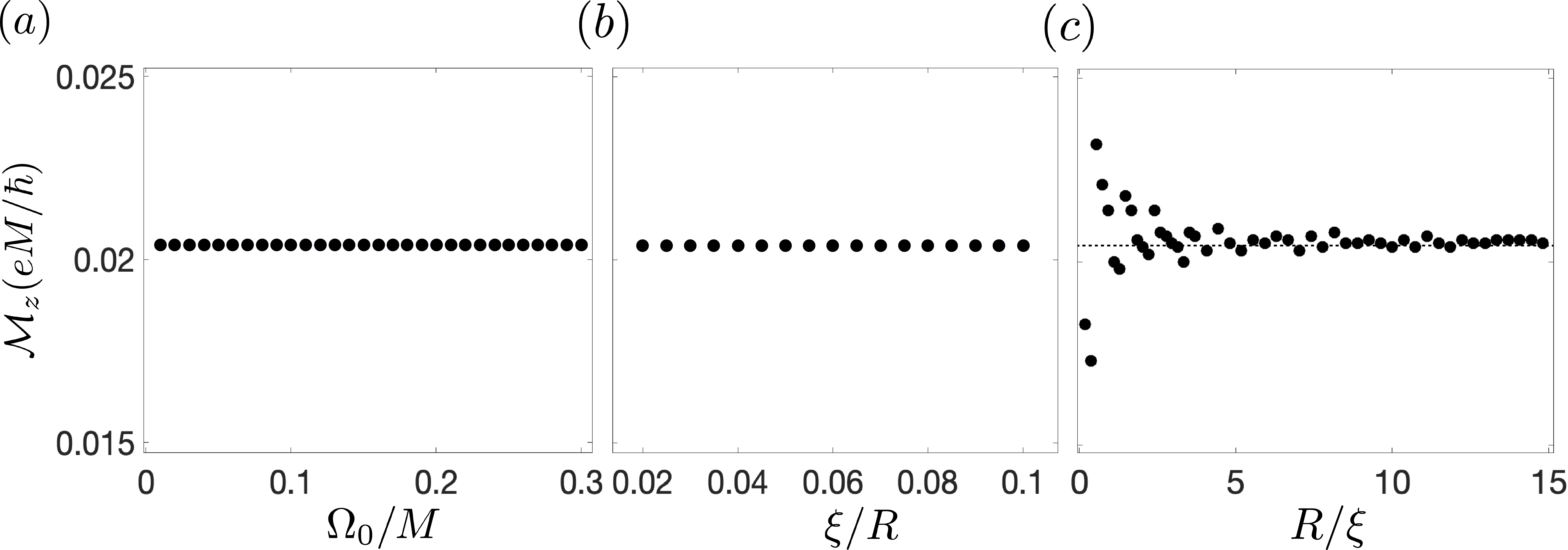}
	\caption{The independence of orbital magnetization density as a function of (a) Rabi frequency, $\Omega_{0}$, (b) light width,  $\xi$, with constant system size $R$, and (c) disc radius $R$, where the light width $\xi$ is constant. Same parameters as in \Cref{F2} are used here, except the one is changed in this figure.}\label{F85}
\end{figure}

\begin{equation}
    \mathbf{m} = -\frac{e}{2}\sum_{\epsilon_{i}<\mu}\bra{\psi_{i}}\mathbf{r}\times\mathbf{v}\ket{\psi_{i}},
\end{equation}
where summation is on occupied states, $\mathbf{v}=-\frac{i}{\hbar}[\mathbf{r},H]$ and the disc area $S=\pi R^{2}$ for $\mathbf{r}=(\hat{x},\hat{y},\hat{z})$. Since our low momentum Floquet theory can capture the physics only around the early Dirac point in the semiconductor, we always set chemical potential $\mu=0$ in calculating the orbital magnetization. One can write the following expression

\begin{eqnarray}\label{orbmag}
&&\mathcal{M}_{z} = \frac{ie}{2S\hbar}\sum_{\epsilon_{i}<\mu} \bra{\psi_{i}}(\hat{x}[\hat{y},H] -\hat{y}[\hat{x},H])\ket{\psi_{i}}.
\end{eqnarray}

The averaged magnetization density $\mathcal{M}_{z}$ can then be defined as the magnetic moment $\mathbf{m}$ per unit area for a 2D system along the z-direction. As it is shown in \Cref{F6}, the averaged magnetization density increases linearly as a function of light vorticity $m$ for both LP and CP cases. Therefore, the orbital magnetization density reaches zero for $m=0$ for the LP light. However, there is a remaining magnetization of CP light for vorticity $m=0$ that results from states hosting circular current density and nonzero magnetic moments. We note that the circular current density can also be observed among bulk and edge states in the case of CP light. Additionally, the averaged orbital magnetization density for the system is independent of the Rabi frequency of light $\Omega_{0}$, the width of light carrying OAM $\xi$, and disc's radius $R$ (intensive quantity) as demonstrated in \Cref{F85}. It should be noted that although in \Cref{F85}(c) the whole range for $R$ is shown, but small disc radius in the range $R \lesssim 5\xi$ is not physical. We note that the independence of the magnetization results from the Rabi frequency may at first seem counter-intuitive especially from the point of view of a driven two-level system. However, upon further scrutiny, it turns out that such a behavior is acceptable within our model. The underlying reason is that in evaluating the magnetization, to simplify our calculations, we have assumed a nearly zero-temperature occupation of the Floquet bands corresponding to a full occupation of the Floquet valence band \cite{seetharam2015controlled}. %which in order to be realized experimentally needs special system-bath engineering techniques. 
However, in a typical experimental setting the dissipative electron-phonon interactions tend to relax the distribution of the electrons from the undriven conduction band to the valence band, so that the occupation probability of electrons in the Floquet valence band around the resonance surface could be significantly lower than one and is controlled by the Rabi frequency \cite{Dehghani2014Dissipative, PhysRevBDehghani2016}. Hence, since in our model we assume that the occupation of the bands is insensitive to the Rabi frequency, the resulting magnetization tends to be independent of the Rabi frequency. 

From a semiclassical point of view, electronic magnetization is determined by the angular speed of electrons. Therefore, based on our results, we can deduce that the effective angular speed of electrons in our system is proportional to the light's vorticity, and is independent of the Rabi frequency and width of the light. For the same semiconductor parameters described in the last paragraph of section \ref{dyncond} and the sample  radius of $R=10\mu m$, the typical evaluated magnetization $\mathcal{M}_z = 0.1 (eM/\hbar)$ yield a total magnetic moment of $\mu = \mathcal{M}_{z}\times \pi R^{2} \simeq 10^{9}\mu_{B}$ in terms of Bohr magneton. Such a magnetic moment can be probed by sensitive SQUID scanning microscopy measurements \cite{AnnalsSQUIDS2022Persky}.
The nonzero magnetization would indicate the existence of nonzero current densities as some examples are calculated in \Cref{F5}. To measure the magnetization spatially with a nano-scale resolution, one can use the magnetometry based on nitrogen vacancies (NV) center in the diamond \cite{hong2013nanoscale,glenn2018high,thiel2019probing,sun2021magnetic}. However, since our Floquet system can be only realized transiently, ultrafast measurement devices which can measure transient magnetic signals is required. We also note that energy of vortex states can be visible by angle-resolved photoemission spectroscopy (ARPES) measurements as this method can acquire the energy dispersion of the system \cite{wang2013observation,mahmood2016selective}.

Regarding the applicability of a low-momentum treatment of the Hall conductivity and the magnetization, we note that because of the structured profile of the light, the translational symmetry is broken and we only can sum over states that are being created from our low momentum theory. Therefore, our results for the magnetization depend on the pseudo-OAM cutoff of the dispersion, ($l_{c}$), albeit weakly. As a result, the magnetization calculated here can represent only the order of magnitude of the magnetization in the Floquet system instead of its exact values. However, since generally, in the presence of translational symmetry, the magnetization density depends on the momentum derivatives of the wave functions \cite{Fukuyama1971}, we expect that for our system, the main contribution to the magnetization density should be attributable to the region in the vicinity of the Dirac points where the curvature of the bands is significant, and we expect it to be captured by our low-momemntum theory. In particular, in \cite{PhysRevB2016Guti} it is demonstrated that in a gapped graphene system, only momenta around the Dirac points K and K$'$ contribute to the magnetization. Therefore, this assumption that our low momentum theory is calculating the main part of the magnetization should be valid.

\section{Discussion and outlook}\label{conclusion}
In this study, two physical observables -- optical conductivity and orbital magnetization -- of the Floquet system driven by a structured light carrying nonzero OAM are calculated. While we only considered the modification of the electronic band structure from the OAM light beam, it is a stepping stone to adding electronic interactions in the system which may realize novel many-body states. In particular, the possibility of creating exotic states in the presence of non-equilibrium superconducting phases in semimetals, semiconductors, and strongly correlated materials \cite{Dehghani2017Dynamical,Claassen2018Universal, Dehghani2021Light, Kennes2019Light,  Dehghani2020Optical, kitamura2021floquet} could be the subject of future research.

\section*{Acknowledgments}
We thank Julia Sell and Bin Cao for helpful discussions. This work is supported by ARL W911NF1920181, AFOSR MURI FA9550-19-1-0399, AFOSR 95502010223, NSF DMR-2019444, ARO W911NF2010232, Minta Martin and Simons Foundations. M.H. thanks ETH Zurich for their hospitality during the conclusion of this work. 

\bibliography{OptHall}

\appendix
\section{Rotating wave approximation: Case of linearly polarized light}\label{appA}
In this appendix, we review the numerical details of the diagonalizing the RWA Hamiltonian for the Floquet system driven by LP light carrying OAM \cite{PhysRevBHwanmun}. In Appendix \ref{CP_RWA}, the procedure is discussed for the case of CP light. The rotating wave approximated Hamiltonian reads as follows

\begin{eqnarray}\label{H_rwa}
H_{\text{RWA}} &=& \frac{v^2}{2M}\left(\mathbf{k}^2 - k_0^2\right)\sigma_z + \left[evA_0(r) e^{-im\phi} \sigma_+ + \text{H.c.}\right] \nonumber\\
&& + O\left(evA_\text{max}\frac{vk_0}{M}\right) \nonumber\\
&=& \frac{\delta}{2}\left(\frac{\mathbf{k}^2}{k_0^2} - 1\right)\sigma_z + \left[ \Omega(r) e^{-im\phi} \sigma_+ + \text{H.c.} \right] \nonumber\\
&& + O\left(\Omega_0\sqrt{\frac{\delta}{M}}\right),
\end{eqnarray}
where the wavevector $\mathbf{k}$ is replaced with the momentum operator $\hat{\mathbf{k}} = (-i\partial_x, -i\partial_y) $.

Due to the commutativity of the effective Hamiltonian given in \eqnref{H_rwa} and the electronic pseudo-OAM $\hat{l}=-i\partial_\phi+(m/2+1)\sigma_z$, $l$ is a good quantum number. To show this, we can write $[-i\partial_\phi,k_x]=ik_y$ and $[-i\partial_\phi,k_y]=-ik_x$. This results in $[-i\partial_\phi,k_x\pm ik_y]=\pm(k_x\pm ik_y)$ and $[-i\partial_\phi,\mathbf{k}^2]=0$. Then
\begin{eqnarray}
\left[-i\partial_\phi, H_{\text{RWA}}\right] &=& -m\left(\Omega(r) e^{-im\phi }\sigma_+ -\text{H.c.} \right), \nonumber\\
\left[\sigma_z, H_{\text{RWA}}\right] &=& 2\left( \Omega(r) e^{-im\phi} \sigma_+ -\text{H.c.} \right),
\end{eqnarray}
and we can immediately conclude $[-i\partial_\phi + (m/2)\sigma_z, H_{\text{RWA}}]=0$. Therefore, $l$ is a conserved quantity and we can block-diagonalize $H_{\text{RWA}}$ according to $l$. Eigenstates of the effective Hamiltonian can be acquired in the following form as vortex states
\begin{eqnarray}\label{eigstate}
    \psi_{n,l}(\mathbf{r})=\begin{pmatrix}
    e^{i(l-m/2-1)\phi}u_{+,n,l}(r)\\
    e^{i(l+m/2+1)\phi}u_{-,n,l}(r)
    \end{pmatrix},
\end{eqnarray}
where, $n$ is the band index, and $m$ is the light vorticity which determines the number of vortex states branches. We can reach the following eigenvalue equations in determining the eigenfunctions presented in \eqnref{eigstate}. The eigenstates satisfy
\begin{equation}
	\begin{split}\label{heff}
E_{n,l} u_{\pm,n,l}(r) &= \mp \frac{\delta^2}{2k_0^2}
\bigg( \partial_r^2 + \frac{1}{r}\partial_r -\\
&\frac{(l\mp m/2)^2}{r^2} + k_0^2\bigg) u_{\pm,n,l}(r)
 + \Omega(r) u_{\mp,n,l}(r).
\end{split}
\end{equation}
By changing $l$ with $-l$ here, we can show that $\psi_{n,-l}(\mathbf{r})=i\sigma_y \psi_{|m|+1-n,l}^*(\mathbf{r})$ and $E_{n,-l}=-E_{|m|+1-n,l}$. Next, we diagonalize the Hamiltonian numerically. We diagonalize the Hamiltonian $h(\mathbf{k})=(M/v^{2})H_{RWA}(\mathbf{k})$ based on the basis functions $\lbrace u_{\pm,n}(r)\rbrace$ such that
\begin{eqnarray}\label{besseleqn}
\left[\partial_r^2 + \frac{1}{r}\partial_r - \frac{l_\pm^2}{r^2} + k_0^2\pm 2\epsilon_{\pm,\alpha}\right]u_{\pm,\alpha}(r)=0,
\end{eqnarray}
where $l_\pm = l\mp(m/2+1)$. \eqnref{besseleqn} is the Bessel's differential equation and we simply find that $u_{\pm,n}(r)=C_{\pm,n} J_{l_\pm}(\sqrt{k_{0}^{2}\pm2\epsilon_{\pm,n}}r)$. We note that $u_{\pm,n}(r)$ are confined on the disc of radius $R$, obeying the boundary condition $u_{\pm,n}(R)=0$ where eigenenergies $\epsilon_{\pm,n}$ are set to be bounded at $r=0$. $C_{\pm,n}$ are normalization constants and determined by the polar coordinate integral condition $\int_0^R |u_{\pm,n}(r)|^2 r dr=1$. Then we have
\begin{eqnarray}
\sqrt{(k_0^2\pm 2\epsilon_{\pm,\alpha})}R=z_n^{(l_\pm)},
\end{eqnarray}
where $z^{(\nu)}_n$ is the $n$th non-negative zero of the Bessel function of order $\nu$, $J_\nu(z)$. We take eigenfunctions near zero energy with the $N$-smallest positive eigenenergies and the $N$-largest negative eigenenergies for each $u_{\pm,n}(r)$ among all infinite possible eigenfunctions $u_{\pm,n}(r)$. Now we can calculate the Hamiltonian components as follows
\begin{eqnarray}
&& M_{s,s'}=\int_0^\infty  u_{+,i_0+s}(r) \Omega(r) u_{-,j_0+s'}(r) r dr, \nonumber
\end{eqnarray}
where we label such eigenfunctions as $n=i_0+1,\cdots,i_0+2N$ for $u_{+,n}(r)$ and $n=j_0+1,\cdots,j_0+2N$ for $u_{-,n}(r)$. The block-diagonal components have the form $(H_+)_{s,s'} = v^2 \epsilon_{+,i_0+s}\delta_{s,s'}/M$ and $(H_-)_{s,s'} = v^2 \epsilon_{-,j_0+s}\delta_{s,s'}/M$, we finally have
\begin{eqnarray}\label{Hamnum}
H_\text{eff,proj}^{(l)}=
\left( \begin{array}{cc} H_+ & M \\ M^\dag & H_- \end{array} \right),
\end{eqnarray}
and we can diagonalize this $4N\times4N$ matrix to obtain the low-energy spectrum and wavefunctions. We note that since eigenstates of the Hamiltonian in \eqnref{Hamnum} are obtained in the Bessel function basis $u^{\pm}$, in order to reconstruct the eigenfunctions in the real space, we should calculate the linear combination of $u^{\pm}$ with corresponding coefficients obtained from Hamiltonian eigenstates. 

\section{Case of circularly polarized light}\label{CP_RWA}
Here, we follow a similar approach to diagonalize the RWA Hamiltonian for the CP light \cite{PhysRevBHwanmun}. The vector potential for CP light $\mathcal{A}(\mathbf{r},t) = A(r)e^{i(m\phi+\omega t)}(\mathbf{\hat{x}}+i\mathbf{\hat{y}}) + c.c.$ yields to the following RWA Hamiltonian

\begin{eqnarray}\label{Hrwa_cp}
H_{\text{RWA}} &=& -\frac{ev^3}{2M^2}\left[ (k_x + ik_y)A_0(r) e^{im\phi} (k_x + ik_y) \sigma_- + \text{H.c.} \right] \nonumber\\
&& + \frac{v^2}{2M}(\mathbf{k}^2 - k_0^2)\sigma_z + O\left(\frac{v^3 k_0^3}{M^2}\right) \nonumber\\
&=& -\frac{\delta}{2M}\left[ \frac{(k_x + ik_y)}{k_0}\Omega(r) e^{im\phi} \frac{(k_x + ik_y)}{k_0} \sigma_- + \text{H.c.} \right] \nonumber\\
&& + \frac{\delta}{2}\left(\frac{\mathbf{k}^2}{k_0^2} - 1\right)\sigma_z + O\left( \delta \sqrt{\frac{\delta}{M}} \right).
\end{eqnarray}

We note that the CP laser field makes the system topological Floquet insulator with Chern number two far from the center of the light. As a result, the system has edge states that are localized at the boundary of the shining light. Similar to the LP light, it can be shown that the pseudo-OAM $\hat{l}=-i\partial_\phi+(m/2+1)\sigma_z$ is a good quantum number and the dispersion can be calculated in terms of $l$. As it was shown for the linear polarization laser field, we use $[-i\partial_\phi,k_x]=ik_y$, $[-i\partial_\phi,k_y]=-ik_x$, $[-i\partial_\phi,k_x\pm ik_u]=\pm(k_x\pm ik_y)$, and $[-i\partial_\phi,\mathbf{k}^2]=0$, therefore

\begin{eqnarray}
&&[-i\partial_\phi,H_{RWA}] \\
&& = -(m+2)\frac{v^{2}\Omega(r)}{2M^{2}}\left[(k_x-ik_y)e^{-im\phi}(k_x-ik_y)\sigma_+ -\text{H.c.} \right], \nonumber\\
&& [\sigma_z,H_{RWA}] =\frac{v^{2}\Omega(r)}{M^{2}}\left[(k_x-ik_y)e^{-im\phi}(k_x-ik_y)\sigma_+ -\text{H.c.} \right], \nonumber
\end{eqnarray}
that results in $[-i\partial_\phi+(m/2+1)\sigma_z,h]=0$. Therefore, we block diagonalize the Hamiltonian $H_{RWA}$ along $l$. The general form of the eigenstates of $H_{RWA}$ are as follows
\begin{eqnarray}
\psi_l(\mathbf{r})=\left( e^{il_+\phi}u^{+}(r) , e^{il_-\phi}u^{-}(r) \right)^T, \qquad
\end{eqnarray}
where $l_\pm= l\mp(m/2+1)$. With this form of wave functions, the eigenvalue equation in the polar coordinate reads as follows 

\begin{eqnarray}\label{heff2}
\left( \epsilon +\frac{\beta}{2r^2} \right)u^{+}(r)
&&= -\frac{1}{2}\left(\partial_r^2 +\frac{1}{r}\partial_r -\frac{\alpha^2}{r^2} + k_0^2 \right)u^{+}(r) \nonumber\\
&&+\frac{\Omega(r)}{2M}\left( \partial_r^2 + \frac{2l+1}{r}\partial_r + \frac{l_+ l_-}{r^2} \right)u^{-}(r) \nonumber\\
&& +\frac{\Omega'(r)}{2M}\left( \partial_r + \frac{l_-}{r} \right) u^{-}(r), \nonumber\\
\left( \epsilon +\frac{\beta}{2r^2} \right)u^{-}(r) &&= \frac{1}{2}\left(\partial_r^2 +\frac{1}{r}\partial_r -\frac{\alpha^2}{r^2} + k_0^2 \right)u^{-}(r) \nonumber\\
&&+\frac{\Omega(r)}{2M}\left( \partial_r^2 -\frac{2l-1}{r}\partial_r + \frac{l_+ l_-}{r^2} \right)u^{+}(r) \nonumber\\
&& +\frac{\Omega'(r)}{2M}\left( \partial_r - \frac{l_+}{r} \right) u^{+}(r),
\end{eqnarray}
where $\alpha=\sqrt{l^2+(m/2+1)^2}$ and $\beta=l(m+2)$. Similar to the case of LP light, we assume that the system is finite size on a disc of radius $R$. We diagonalize the Hamiltonian in the basis functions $\{u_{\pm,\alpha}(r)\}$ satisfying \eqnref{besseleqn}. We have the similar boundary condition $u_{\pm,\alpha}(R)=0$ for $\alpha \in \mathbb{N}$ that yields to Bessel functions solutions and normalization for \eqnref{besseleqn} as it was discussed in the main text. Therefore, we have

\begin{eqnarray}
\epsilon_{\pm,\alpha}=\pm\frac{1}{2}\left(\frac{z^{(l_\pm)}_\alpha}{R}\right)^2\mp\frac{k_0^2}{2}.\ \qquad
\end{eqnarray}

Here, $z_{\alpha}^{(l_{\pm})}$ is the $\alpha$th zero of Bessel function with order $l_{\pm}$. Again, we can label the eigenfunctions as $\alpha=i_0+1,\cdots,i_0+2N$ for $u_{+,\alpha}(r)$ and $\alpha=j_0+1,\cdots,j_0+2N$ for $u_{-,\alpha}(r)$ as we have truncated the Hamiltonian for the $N$ eigenfunctions below and $N$ eigenfunctions above the zero energy for the basis $\{u_{\pm,\alpha}(r)\}$, the same as \eqnref{Hamnum}. Now we can construct the effective projected Hamiltonian by components of the following form

\begin{eqnarray}
&& M_{s,s'}=\int_0^\infty \left[ u_{+,i_0+s}(r)\frac{v^{2}\Omega(r)}{k_0^2}
\left\lbrace \partial_r^2 u_{-,j_0+s'}(r) \frac{}{} \right. \right. \\
&& \ \left. +\frac{2l+1}{r}\partial_r u_{-,j_0+s'}(r) +\frac{l_+ l_-}{r^2} u_{-,j_0+s'}(r) \right\rbrace \nonumber\\
&& \ \left. + u_{+,i_0+s}(r)\frac{v^{2}\Omega'(r)}{k_0^2}\left\lbrace \partial_r u_{-,j_0+s'}(r) +\frac{l_-}{r} u_{-,j_0+s'}(r)\right\rbrace\right] r dr. \nonumber
\end{eqnarray}

Then, we can similarly build the Hamiltonian defined in \eqnref{Hamnum}, with block-diagonal terms of the form $(H_+)_{s,s'} = v^2 \epsilon_{+,i_0+s}\delta_{s,s'}/M$ and $(H_-)_{s,s'} = v^2 \epsilon_{-,j_0+s}\delta_{s,s'}/M$. After diagonalizing the Hamiltonian, one can acquire the dispersion as depicted in \Cref{F2}(b). Similar to the case of LP light, there are $|m|$ vortex state branches. 

\section{Dynamical conductivities for the case of $m=2$}\label{m2LPHall}

In this part, the optical conductivity as a function of the probe field frequency for the LP light with OAM $m=2$ is presented. We can calculate the same conductivities for the vorticity $m=2$ similar to the case of $m=1$ as shown in \Cref{F3} and discussed in the main text. However, the V-V transition has more peaks compared to the same transition for $m=1$, because there are two chiral vortex branches with more available states for electron transfer between inter-and intra-vortex branches satisfying the transition rules. Consequently, more transitions are also possible for the V-B electronic transfers. Similar to the light OAM $m=1$, the B-B transition has the most possible transitions and peaks that corresponding probe frequencies locate in the ranges $v^{2}\Omega_{0}<\omega'$ and decays exponentially at higher probe frequencies.

\begin{figure*}[ht]
	\centering
	\includegraphics[width=0.8\linewidth]{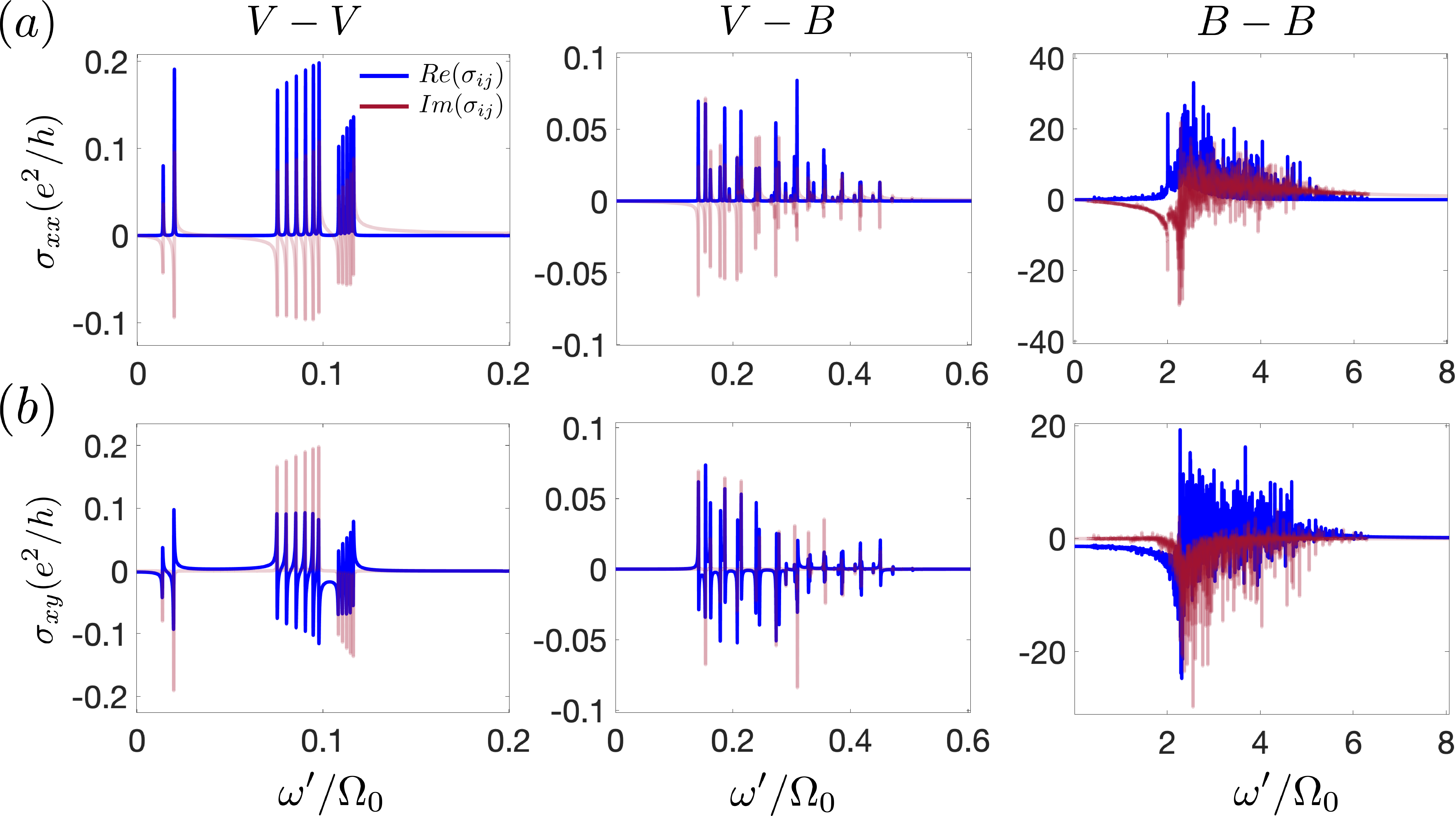}
	\caption{The Hall and longitudinal conductivities of linearly polarized light $ \sigma_{xy} $ and $ \sigma_{xx} $, versus probe field frequency $ \omega'$ and vorticity $m=2$ for contributions arising from different types of transitions. Parameters are the same as in \Cref{F2}. Blue and red lines indicate the real and imaginary parts of optical conductivities, respectively.}\label{F8}
\end{figure*}

\end{document}